\documentclass[twocolumn,english,prl]{revtex4}
\pdfoutput=1
\usepackage{pslatex}
\usepackage[T1]{fontenc}
\usepackage[latin1]{inputenc}
\usepackage{float}
\usepackage{graphicx}
\usepackage{amssymb}

\makeatletter

\providecommand{\tabularnewline}{\\}

\usepackage{babel}
\makeatother
\begin{document}

\title{Observing \emph{Zitterbewegung} with Ultracold Atoms}

\author{J. Y. Vaishnav}

\affiliation{Joint Quantum Institute, National Institute of Standards and Technology,
Gaithersburg MD 20899 USA}

\author{Charles W. Clark}

\affiliation{Joint Quantum Institute, National Institute of Standards and Technology,
Gaithersburg MD 20899 USA}

\begin{abstract}
We propose an optical lattice scheme which would permit the experimental
observation of Zitterbewegung (ZB) with ultracold, neutral atoms.
A four-level ``tripod'' variant of the usual setup for stimulated
Raman adiabatic passage (STIRAP) has been proposed for generating
non-Abelian gauge fields \citep{ruseckas-2005-95}. Dirac-like Hamiltonians,
which exhibit ZB, are simple examples of such non-Abelian gauge fields;
we show how a variety of them can arise, and how ZB can be observed,
in a tripod system. We predict that the ZB should occur at experimentally
accessible frequencies and amplitudes.
\end{abstract}
\maketitle
A driving force behind the study of ultracold atoms is their potential
use as highly tunable quantum simulators for physical systems, ranging
from quantum phase transitions in solids \citep{jaksch1998cba} to
black holes \citep{garay2000sag}. In particular, the high degree
of control over length and time scales in cold atom experiments allows
for the possibility of observing phenomena that are experimentally
inaccessible in their original counterpart systems. In this paper,
we propose an experiment which simulates the relativistic (and recently,
controversial \citep{krekora2004rel}) phenomenon of \emph{zitterbewegung}
(ZB), a jittering motion caused by interference between the positive
and negative energy components of the wavefunction of a Dirac fermion.

For a relativistic electron, the ZB frequency is of the order of $mc^{2}/\hbar\approx10^{20}$
$\mbox{s}^{-1}$, and the amplitude comparable to the Compton wavelength,
$h/mc\approx10^{-12}\mbox{ m}$. ZB has never been observed for free
electrons, as these time and length scales render it experimentally
inaccessible. The presence of ZB is, however, a general feature of
spinor systems with linear dispersion relations. Trapped ions \citep{lamata2007dea}
as well as condensed matter systems, including graphene \citep{katsnelson2007cqg,cserti2006udz,rusin-2007}
and semiconductor quantum wires \citep{schliemann2005zew,schliemann2006zea},
have been proposed as candidate systems for observing ZB.

In this paper, we propose a scheme for observing ZB in ultracold neutral
atoms. A four-level {}``tripod'' variant of the usual setup for
stimulated Raman adiabatic passage (STIRAP) has previously been proposed
for generating non-Abelian gauge fields \citep{ruseckas-2005-95}.
Dirac-like Hamiltonians, which exhibit ZB, are simple examples of
such non-Abelian gauge fields, and we show how a variety of them can
arise in a tripod system. The Hamiltonian for atoms in an optical
lattice is Dirac-like in the subspace of the tripod's two degenerate
dark states. We predict that an atom's mean position should thus undergo
Dirac-like ZB. However, the characteristic amplitude of tripod ZB
is the optical lattice wavelength, vs. the Compton wavelength of Dirac
ZB, and the oscillation energy is proportional to the lattice recoil
energy vs. the rest mass energy of the electron. This places tripod
ZB well within the range of experimental observation, with a characteristic
frequency of MHz vs. the THz domain predicted for condensed matter
implementations \citep{rusin-2007}. Although we treat here the case
of a noninteracting gas, the ZB persists under the addition of weak
interactions: The Hamiltonian separates into center of mass and relative
coordinates, and the center of mass Hamiltonian is again Dirac-like.
In a dilute atomic cloud, the ZB should thus manifest itself as an
oscillation of the cloud's center of mass.

\begin{figure}[!tbh]
\begin{centering}
\includegraphics[width=0.65\columnwidth]{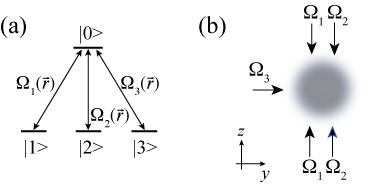}
\par\end{centering}

\caption{Left: Tripod STIRAP scheme with Rabi frequencies as defined in Eqs.
(\ref{eq:rabi1}-\ref{eq:rabi3}). Right: Schematic geometry of tripod
laser beams yielding the Rabi frequencies of Eqs. (\ref{eq:rabi1}-\ref{eq:rabi3}).
\label{fig:tripod}}
\end{figure}

We consider the tripod STIRAP scheme described in \citep{ruseckas-2005-95}
and shown in Fig. \ref{fig:tripod}. %
The Hamiltonian in the interaction picture is $H=-\hbar\sum_{i=1}^{3}\Omega_{i}|0\rangle\langle i|+\mbox{h.c.}$
Defining $\Theta_{i}=\sqrt{\sum_{j=1}^{i}\left|\Omega_{j}\right|^{2}}$,
the dressed states include two dark states degenerate at zero energy:
$|D_{1}\rangle=\frac{1}{\Theta_{2}}\left(\Omega_{2}|1\rangle-\Omega_{1}|2\rangle\right)$
and $|D_{2}\rangle=\frac{1}{\Theta_{3}}\left(\Omega_{1}^{*}\Omega_{3}|1\rangle-\Omega_{2}^{*}\Omega_{3}|2\rangle-\Theta_{2}^{2}|3\rangle\right)$
(we have chosen an orthonormal basis). Suppose the atoms are now slowly
moving in the field. The degeneracy causes the Born-Oppenheimer approximation
to break, yielding an effective U(2) non-Abelian gauge field. The
effective Hamiltonian in the $2\times2$ dark subspace is $H=\frac{1}{2m}(\vec{p}-\mathbf{\hat{A}})^{2}+\hat{\Phi}$
where $m$ is the atom's mass, $\mathbf{A}_{i,j}=i\hbar\langle D_{i}|\vec{\nabla}|D_{j}\rangle$
is an effective vector potential, and $\hat{\Phi}$ is a scalar Born-Huang
potential resulting from the coupling to the bright subspace. The
following choice of Rabi frequencies\begin{eqnarray}
\Omega_{1}(\vec{r}) & = & \Omega\sqrt{1-\epsilon^{2}}\cos k_{0}z\label{eq:rabi1}\\
\Omega_{2}(\vec{r}) & = & \Omega\sqrt{1-\epsilon^{2}}\sin(k_{0}z+\pi)\label{eq:rabi2}\\
\Omega_{3}(\vec{r}) & = & \epsilon\Omega e^{ik_{0}y},\label{eq:rabi3}\end{eqnarray}
for $\Omega$, $\epsilon,$ and $k_{0}$ constant corresponds to the
laser beam geometry in Fig. \ref{fig:tripod}(b), and yields a Dirac-like
Hamiltonian (a related setup was proposed in \citep{galitski} in
the context of observing spin relaxation effects). Specifically, after
some trivial gauge transformations, the two dark states feel an effective
vector potential\begin{eqnarray}
\hat{A}_{y} & = & \frac{\hbar k_{0}}{2}\left(1-\epsilon^{2}\right)\sigma_{z}\label{eq:ayfree}\\
\hat{A}_{z} & = & -\epsilon\hbar k_{0}\sigma_{y}\label{eq:azfree}\end{eqnarray}
and an effective scalar potential $\hat{\Phi}=V_{0}\sigma_{z},$ where
$V_{0}=\frac{\hbar^{2}k_{0}^{2}}{2m}\left(1-\epsilon^{2}\right).$
In general, $[\hat{A}_{y},\hat{A}_{z}]\neq0,$ and the field is non-Abelian.
The vector potential in Eqs. (\ref{eq:ayfree}-\ref{eq:azfree}) is
valid for $-1<\epsilon<1$, acquiring an extra normalization otherwise.

We can write the full Hamiltonian in the dark subpace as \begin{equation}
H=\frac{p^{2}}{2m}-\frac{\hbar k_{0}}{2m}\left[(1-\epsilon^{2})p_{y}\sigma_{z}-2\epsilon p_{z}\sigma_{y}\right]+V_{0}\sigma_{z};\label{eq:spinorbit}\end{equation}
we shall henceforth not consider the free $x$ direction. Thinking
of $|D_{1(2)}\rangle$ as {}``spin up (down),'' the Hamiltonian
of Eq. (\ref{eq:spinorbit}) is effectively a spin-orbit coupling,
in the presence of a homogeneous magnetic field along the $z$ direction.
In fact, letting $\alpha=\frac{\hbar k_{0}}{m}\epsilon,$ defining
$\epsilon_{R}=-1+\sqrt{2},$ and choosing $\epsilon=\epsilon_{R}$,
we have $H=\frac{p^{2}}{2m}+\alpha(p_{y}\sigma_{z}-p_{z}\sigma_{y})+V_{0}\sigma_{z}$--the
Hamiltonian for a 2D electron gas in the $y$--$z$ plane, with Rashba
spin-orbit coupling, and subject to a homogeneous magnetic field along
the $z$ axis. Ref. \citep{dudarev-2004-92} proposes an alternate
scheme for generating spin-orbit coupling with ultracold atoms.

It is possible to remove the scalar potential by applying a state-dependent
external potential to the system. Denoting the potential felt by $|i\rangle$
as $V_{i}(\vec{r})$, choosing $V_{1}(\vec{r})=V_{2}(\vec{r})=V(\vec{r})$,
and $V_{3}(\vec{r})=V(\vec{r})+V_{0}(1+\epsilon^{2})/(1-\epsilon^{2})$
subjects the dark states to an additional potential $\hat{V}=V(\vec{r})\otimes\mathbf{I}-\hat{\Phi}.$
If the scalar potential is thus removed from Eq. (\ref{eq:spinorbit}),
the resulting Hamiltonian is extremely versatile, for two reasons:
(1) $\epsilon$ is tunable and (2) the dark states form a degenerate
subspace, for which any basis is equivalent. In fact, depending on
the direction we call {}``spin up,'' this Hamiltonian can be viewed
as a variety of Dirac-like Hamiltonians (see Table \ref{dirac}).

\begin{table*}
\noindent \begin{centering}
\begin{tabular}{|c|c|c|c|}
\hline 
{}``Spin-up''&
$\epsilon$&
$H$&
Analog\tabularnewline
\hline
\hline 
$|D_{1}\rangle$&
$\epsilon_{R}$&
$H=\frac{p^{2}}{2m}+\alpha(p_{y}\sigma_{z}-p_{z}\sigma_{y})$&
Rashba\tabularnewline
\hline 
$\frac{1}{\sqrt{2}}\left(|D_{1}\rangle+i|D_{2}\rangle\right)$&
$\epsilon_{R}$&
$H=\frac{p^{2}}{2m}+\alpha(p_{y}\sigma_{y}-p_{z}\sigma_{z})$&
Linear Dresselhaus\tabularnewline
\hline 
$\frac{1}{\sqrt{2}}\left(|D_{1}\rangle+i|D_{2}\rangle\right)$&
$-\epsilon_{R}$&
$H=\frac{p^{2}}{2m}+\alpha(p_{y}\sigma_{y}+p_{z}\sigma_{z})$&
Graphene sheet, in vicinity of Dirac point\tabularnewline
\hline
\end{tabular}
\par\end{centering}

\caption{By tuning $\epsilon$ and choosing different states to represent
spin-up, the tripod setup can replicate a variety of Dirac-like Hamiltonians
(see e.g. \citep{cserti2006udz}). Ref. \citep{zhu2007sad} proposes
an alternate method of generating a graphene-like Hamiltonian. \label{dirac}}
\end{table*}

The eigenstates of the system are spinors of the form $e^{ik_{y}y}e^{ik_{z}z}\otimes$$|i;\vec{k}\rangle$
with $i=a,b$. The dispersion relation of the Hamiltonian in Eq. (\ref{eq:spinorbit})
is\begin{equation}
E_{\pm}(k_{y},k_{z})=\frac{\hbar^{2}k^{2}}{2m}\pm\frac{\hbar^{2}k_{0}}{m}\sqrt{(1-\epsilon^{2})^{2}{(k}_{y}-k_{0})^{2}+(2\epsilon)^{2}k_{z}^{2}.}\label{eq:dispersionci}\end{equation}
\begin{figure}[!tbh]
\begin{centering}
\includegraphics[width=0.6\columnwidth]{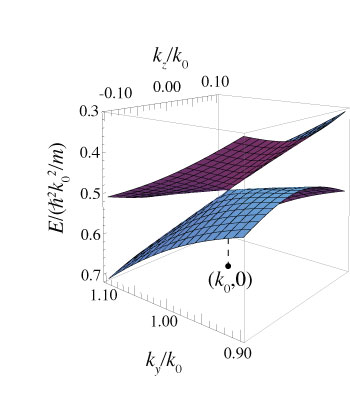}
\par\end{centering}

\caption{(Color online) Energy surfaces in momentum space for $^{85}\mbox{Rb}$
with $k_{0}=2\pi/820\,\mbox{nm}^{-1}$ and $\epsilon=\epsilon_{R}.$
A conical intersection occurs at ${(k}_{y},k_{z})={(k}_{0},0)$; a
circuit of this conical intersection in $\vec{k}$ space gives a Berry
phase.\label{fig:diabolical}}
\end{figure}
 The energy surfaces in Eq. (\ref{eq:dispersionci}) have a conical
intersection at ${(k}_{y},k_{z})={(k}_{0},0)$ (see Fig. \ref{fig:diabolical}).
A circuit of the degeneracy in momentum space yields a Berry phase:
Defining \begin{equation}
\tan\xi(\vec{k})=\frac{2\epsilon k_{z}}{(1-\epsilon^{2}){(k}_{y}-k_{0})},\label{eq:phaseangle}\end{equation}
we designate the eigenfunctions by $|a;\vec{k}\rangle=[i\sin\frac{\xi(\vec{k})}{2},\cos\frac{\xi(\vec{k})}{2}]^{T},$
$|b;\vec{k}\rangle=[i\cos\frac{\xi(\vec{k})}{2},-\sin\frac{\xi(\vec{k})}{2}]^{T}$
which are multivalued for a particular $(k_{y},k_{z}).$ The momentum-space
Berry phase is much like the one encountered in graphene, which gives
rise to phenomena like the half-integer quantum Hall effect \citep{zhang2005eoq}.
We now examine the role that the Berry phase plays in generating ZB.

Consider the time evolution of a Gaussian wavepacket prepared in a
superposition of dark states. In the non-Abelian case ($0<\epsilon<1$),
the eigenvectors have an associated Berry phase, and are both $\vec{k}$
dependent and multi-valued. However, the initial spin state must be
single valued, forcing its expansion coefficients to be $\vec{k}$
dependent (and also multivalued). The presence of the Berry phase
thus translates into a $\vec{k}$ dependence of $\xi(\vec{k})$ in
Eq. (\ref{eq:phaseangle}); we later show that it is this nonvanishing
$\vec{\nabla}_{\vec{k}}\xi(\vec{k})$ that gives rise, directly, to
ZB. 

Considering an initial wavepacket \begin{equation}
\psi(\vec{r};0)=\frac{1}{\sqrt{2}}\int d\vec{k}g(\vec{k};0)e^{i\vec{k}\cdot\vec{r}}\left(\begin{array}{c}
1\\
1\end{array}\right),\label{eq:initial}\end{equation}
it is straightforward to show that its time evolution is\begin{eqnarray}
\psi(\vec{r};t) & = & \frac{1}{\sqrt{2}}\int d\vec{k}g(\vec{k};0)e^{i\vec{k}\cdot\vec{r}}e^{-iE_{0}(\vec{k})t/\hbar}\label{eq:wavepacket}\\
 &  & \times\left\{ \cos\left[\omega(\vec{k})t\right]\left(\begin{array}{c}
1\\
1\end{array}\right)-ie^{-i\xi(\vec{k})}\sin\left[\omega(\vec{k})t\right]\left(\begin{array}{c}
1\\
-1\end{array}\right)\right\} .\nonumber \end{eqnarray}
where $\omega(\vec{k})=\frac{1}{2\hbar}(E_{+}(\vec{k)}-E_{-}(\vec{k)})$.

ZB, in the Dirac equation, is an oscillation of the average position
$\langle\vec{r}(t)\rangle.$ The usual method of understanding the
Dirac equation, and related equations is to derive equations of motion
for the Heisenberg operators, and show that they oscillate in time
\citep{cserti2006udz,schliemann2005zew,schliemann2006zea,rusin-2007}.
We instead work in the Schrödinger picture, which makes explicit the
connection to Berry phase. As $[H,p]=0,$ it is convenient to work
in the momentum basis, and calculate

\[
\langle\vec{r}(t)\rangle=i\int d\vec{k}\vec{\phi}^{\dagger}(\vec{k};t)\cdot\vec{\nabla}_{\vec{k}}\vec{\phi}(\vec{k};t)\]
where $\vec{\psi}(\vec{r};t)=\int d\vec{k}\vec{\phi}(\vec{k};t)e^{i\vec{k}\cdot\vec{r}}/(2\pi)$
is the spinor wavefunction. After some algebra, we find\begin{eqnarray*}
\langle\vec{r}(t)\rangle & = & \langle\vec{r}(0)\rangle+\frac{\hbar\langle\vec{k}(0)\rangle}{m}t\\
 &  & +\frac{1}{2}\int d\vec{k}\left|g(\vec{k};0)\right|^{2}\left(\vec{\nabla}_{\vec{k}}\xi(\vec{k})\right)\left[1-\cos2\omega(\vec{k})t\right]\end{eqnarray*}
where the final term, which oscillates in time, is ZB. The amplitude
of the oscillation is proportional to $\vec{\nabla}_{\vec{k}}\xi(\vec{k}).$
We had previously shown that the $\vec{k}$ dependence of $\xi(\vec{k})$
occurs as a direct consequence of the eigenfunctions being multivalued.
The Schrödinger picture thus illuminates what is not evident in the
Heisenberg representation--that the ZB here can be viewed as a measurable
consequence of the momentum-space Berry phase.

We now suggest a possible experimental demonstration of ZB using ultracold
atoms. Suppose an ensemble of atoms is prepared in the vibrational
ground state of a harmonic trap. A Raman pulse with space-dependent
Rabi couplings is applied, as suggested in \citep{galitski}, to put
the atom in a superposition of dark states, after which the trap is
switched off to allow ballistic expansion. The initial wavepacket
can be approximated by a Gaussian function $g(\vec{k};0)=\frac{d}{\sqrt{\pi}}e^{-\frac{1}{2}(\vec{k}-\vec{k}^{(i)})^{2}d^{2}},$
where $d$ is the oscillator length of the trap, and $\vec{k}^{(i)}$
is a momentum boost (which is zero for the case of a stationary trap).
For this wavepacket, the expectation values of $y$ and $z$ oscillate
as\begin{eqnarray}
\left(\begin{array}{c}
\langle y(t)\rangle\\
\langle z(t)\rangle\end{array}\right) & = & \frac{\hbar\vec{k}^{(i)}}{m}t+\frac{d}{2\pi}\int d\vec{k}e^{-(\vec{k}-\vec{k}^{(i)})^{2}d^{2}}\label{eq:rt}\\
 &  & \times\frac{1}{\tilde{k}^{2}}\left[1-\cos2\omega(\vec{k})t\right]\left(\begin{array}{c}
(\epsilon^{2}-1)\tilde{k_{z}}\\
2\epsilon\tilde{k_{y}}\end{array}\right)\nonumber \end{eqnarray}
where we have defined $\tilde{k}_{y}=(1-\epsilon^{2})k_{y},$ $\tilde{k}_{z}=2\epsilon k_{z}$,
and $\tilde{k}=\sqrt{\tilde{k}_{y}^{2}+\tilde{k}_{z}^{2}}.$ Eq. (\ref{eq:rt})
shows that the ZB vanishes in the Abelian cases, $\epsilon=0$ or
$\epsilon=1$.

It is useful to consider the limit $d\rightarrow\infty,$ where $g(\vec{k};0)\rightarrow\delta(\vec{k}-\vec{k}^{(i)})$,
i.e., the initial wavepacket approaches a plane wave. The integrals
in Eqs. (\ref{eq:rt}) become trivial, and we find that\begin{eqnarray*}
\left(\begin{array}{c}
\langle y(t)\rangle\\
\langle z(t)\rangle\end{array}\right) & = & \frac{\hbar\vec{k}^{(i)}}{m}t+\frac{1}{2\left(\tilde{k}^{(i)}\right)^{2}}\left[1-\cos2\omega(\vec{k}^{(i)})t\right]\\
 &  & \times\left(\begin{array}{c}
(\epsilon^{2}-1)\tilde{k_{z}}^{(i)}\\
2\epsilon\tilde{k_{y}}^{(i)}\end{array}\right).\end{eqnarray*}
\[
\]
In the opposite limit, $d\rightarrow0,$ the ZB vanishes, and for
intermediate values the energy spread causes damping, as can be shown
analytically for bilayer graphene \citep{rusin-2007}.

\begin{figure}[!tbh]
\begin{centering}
\includegraphics[clip,width=0.8\columnwidth]{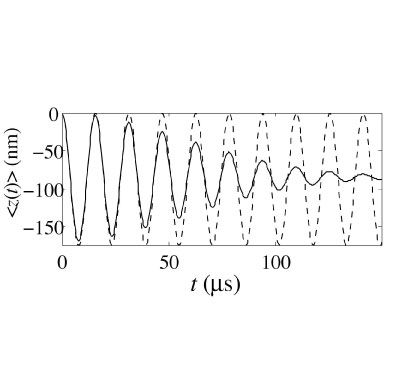}
\par\end{centering}

\caption{ZB for an atom with zero momentum spread (dashed), and for a momentum
spread corresponding to the velocity spread of a cloud initially in
a trap of frequency 112 Hz (solid). ZB oscillations for finite momentum
spread damp out over time, but persist over several periods.\label{fig:zb}}
\end{figure}

Due to the induced Born-Huang field, ZB will occur in this system
(unlike its condensed matter counterparts \citep{rusin-2007,schliemann2005zew,cserti2006udz})
even if the wavepacket has an initial zero group velocity. Supposing
$^{85}$Rb atoms, we take the lattice wavenumber to be $k_{0}=(2\pi/820)\,\mbox{nm}^{-1}$,
and a Gaussian with $\vec{k}^{(i)}=\vec{0}$ and width $k_{0}d=16.2,$
corresponding to the ground state of a trap with trap frequency 112
Hz \citep{suominen1998nde}. Fig. \ref{fig:zb} shows that a pronounced
oscillation would occur in the $z$ direction before damping out.
A typical time scale of the ZB here would be $\mu{\rm s}$ rather
than the ${\rm fs}$ predicted in e.g. graphene and related systems
\citep{rusin-2007}.

\begin{figure}
\begin{centering}
\includegraphics[width=0.9\columnwidth]{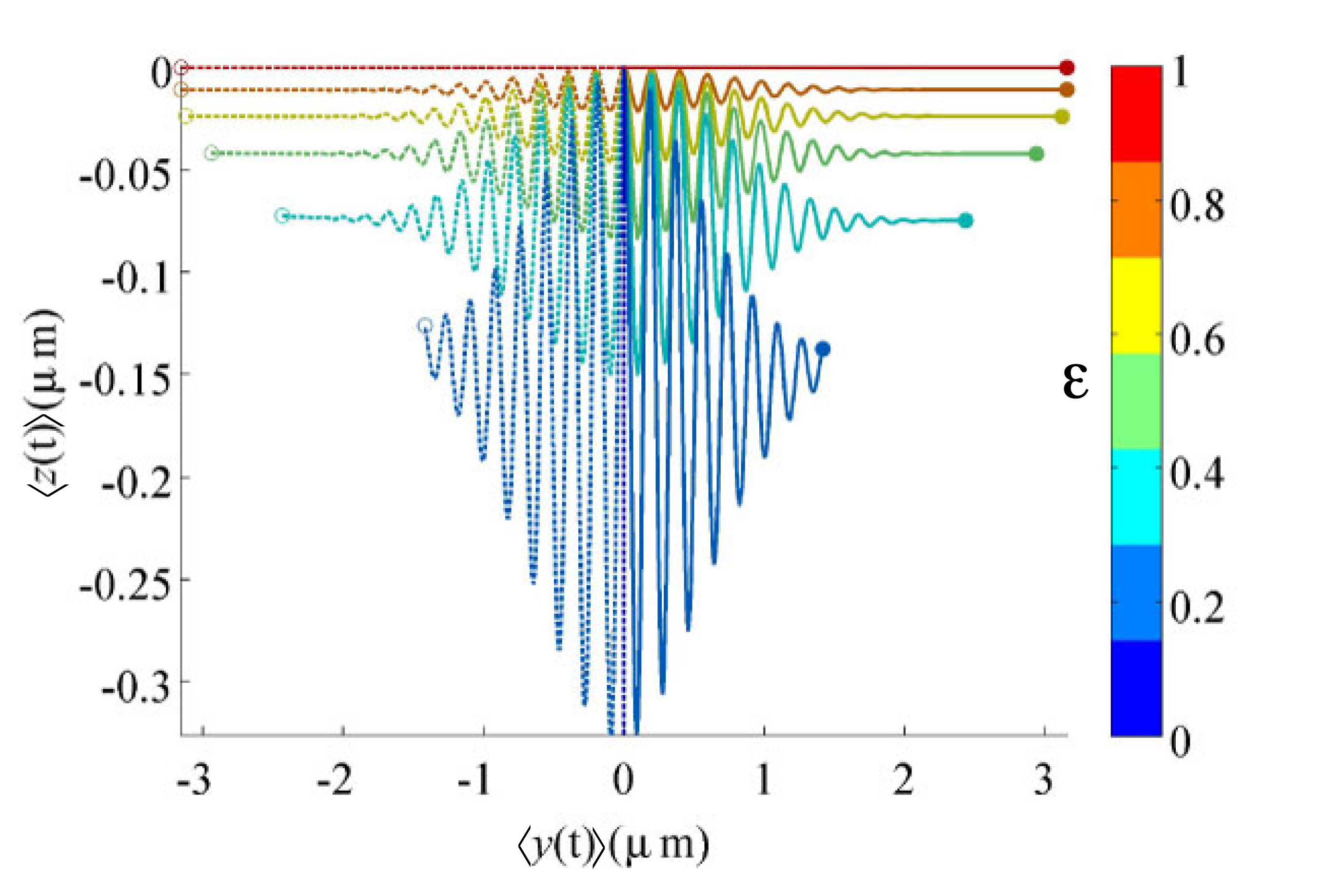}
\par\end{centering}

\caption{(Color online) Spin separation and ZB for $\epsilon=0(0.2)1$, as
indicated on the right colorbar. For each value of $\epsilon$ the
dashed trajectory corresponds to the atom's mean position in {}``spin-up,''
while the solid trajectory corresponds to the mean position of the
atom in {}``spin-down;'' open and closed circles indicate the respective
ends of these trajectories. In the Abelian cases, $\epsilon=0$ and
1, the trajectories are straight lines; the trajectory for $\epsilon=0$
is the vertical line $y=0$. \label{fig:wings}}
\end{figure}

We have shown that the mean position of the atom oscillates sinusoidally.
However, ZB can also be viewed in terms of state-resolved spatial
dynamics. For the Gaussian initial wavepacket, it is not difficult
to show that as the center of mass of the cloud is oscillating in
the $z$ direction, in the $y$ direction, the wavepacket separates
by spin, such that $\langle y_{1,2}(t)\rangle=\pm\hbar k_{0}t/m$
(see Fig. \ref{fig:wings}). This spin separation, which coexists
with the ZB, is a manifestation of the atomic spin Hall effect proposed
in a different setup \citep{spinhall}; a related effect occurs in
velocity-selective coherent population trapping \citep{vscpt}.

\begin{figure}[H]
\begin{centering}
\includegraphics[width=0.7\columnwidth]{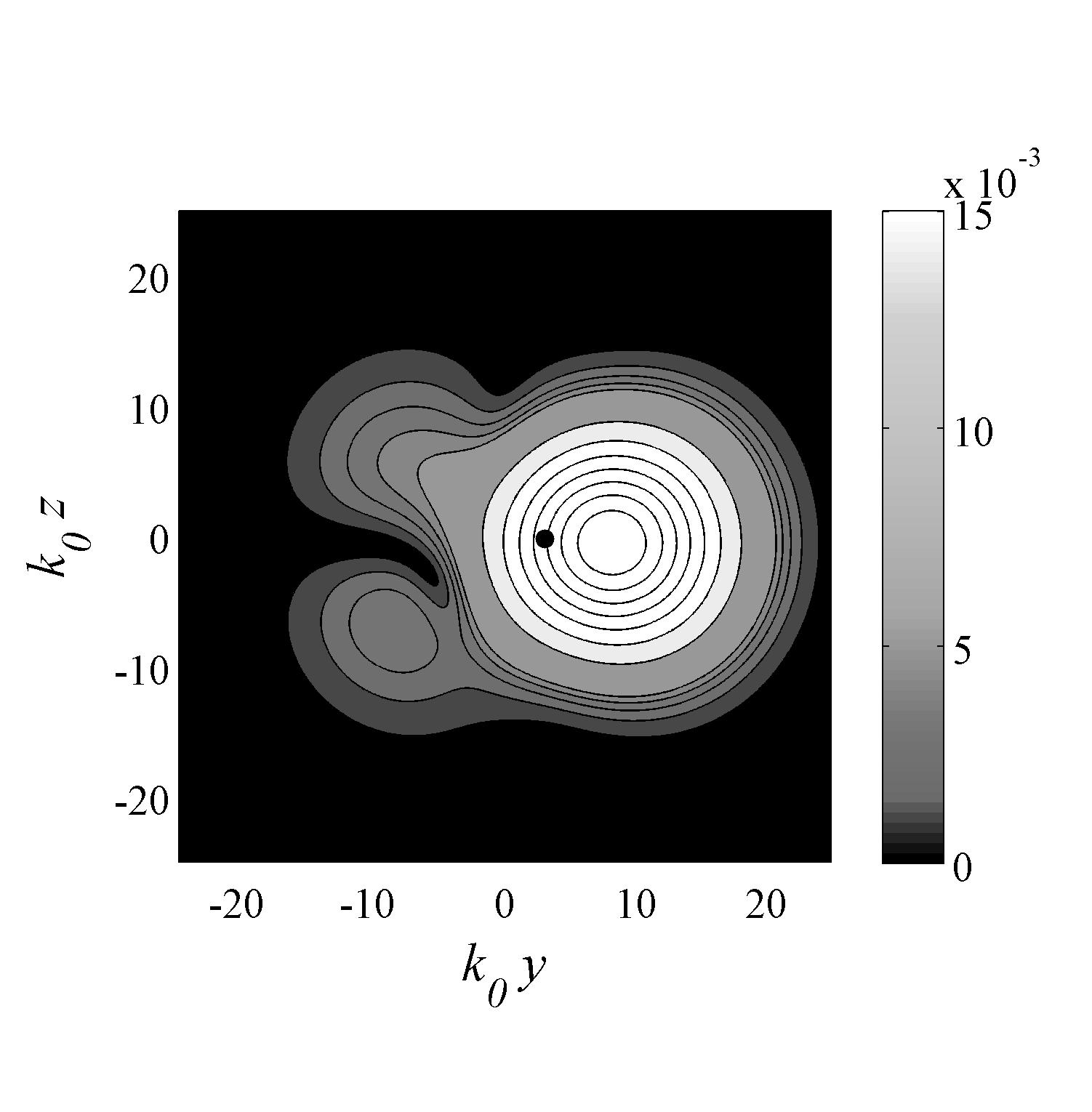}
\par\end{centering}

\caption{As seen in Fig. \ref{fig:wings}, the wavepacket separates by internal
state as it jitters.  This figure shows the time-evolved probability distribution of the "spin-up" component of the wavefunction for an initial Gaussian wavepacket ($\epsilon=\epsilon_{R},$ $k_{0}d=5,$ $\hbar k_{0}^{2}t/m$=10).  The black dot indicates the mean position, which
has drifted to the right. The mixing between internal states gives
rise to ``tails'' on the wavepacket, resulting in ZB. The ZB damps
as the internal states separate. \label{fig:tof}}
\end{figure}

Fig. \ref{fig:tof} shows the dynamics of the {}``spin-up'' component
of the wavepacket in a representative non-Abelian case. In essence,
the effective magnetic field deflects spin-up and spin-down in opposite
directions. As the two wavepackets separate, the coupling between
the components results in oscillating {}``tails'' on the wavepackets,
giving rise to ZB. The ZB decays as the wavepackets separate and cease
to interfere.%

We note that the Hamiltonian for $N$ particles in the non-Abelian
gauge field, interacting via two-body interactions, separates in center-of-mass
and relative coordinates, $\vec{R}$ and $\vec{\rho}$ respectively.
The Hamiltonian is then $\hat{H}=\hat{H}_{CM}+\hat{H}_{\vec{\rho}}$,
where $\hat{H}_{\vec{\rho}}$ is a function only of the relative coordinates
$\vec{\rho}$, and \begin{eqnarray*}
\hat{H}_{CM} & = & \frac{1}{2mN}\vec{P}^{2}-\frac{\hbar k_{0}}{2m}\sum_{i}[(1-\epsilon^{2})(P_{y}+\hbar k_{0})\Sigma_{z}-2\epsilon P_{z}\Sigma_{y}],\end{eqnarray*}
where $\vec{P}$ is the center-of-mass momentum and $\vec{\Sigma}=\sum_{i}\vec{\sigma}^{(i)}$
is the total spin. The Hamiltonian for the center of mass is thus
again of a single-particle Dirac form (with a higher spin), and in
a dilute cloud of $N$ particles with two-body interactions, the center
of mass of the cloud undergoes ZB.

In this paper, we have examined the dynamics of an atom in a tripod
level scheme on an optical lattice; this common experimental setup
gives rise to a non-Abelian gauge field which is isomorphic to the
spin-orbit interaction in 2D electron gases. The idea of ``atomtronics,"
or engineering atomic versions of semiconductor devices, has generated recent
interest \citep{ruschhaupt2004adl,seaman2007aua}.  The prospect of
engineering artificial spin-orbit couplings suggests the possibility
of atomic {}``spintronics,'' for example engineering atomic counterparts
of devices such as the Datta-Das transistor \citep{dattadas}, which
have yet to be successfully realized with electrons. The tripod system
here exhibits atomic ZB, with an amplitude many orders of magnitude
larger than that offered by Dirac electrons or recently discussed
condensed matter systems. We believe it is a promising candidate for
the experimental observation of ZB.

\bibliographystyle{apsrev}

\end{document}